 \documentclass[alpha-refs]{wiley-article}

\usepackage{color}
\usepackage{ulem}

\usepackage{siunitx}
\usepackage{booktabs} 
\usepackage{hyperref} % For hyperlinks in the PDF
\usepackage{amsmath}
\usepackage{lineno}
\usepackage{xr} % able to refer to labels from other documents
\usepackage{pdfpages}

\papertype{Article}
\title{Tropical precipitation clusters as islands on a rough water-vapor topography}

\abbrevs{CWV, column water vapor; KPZ, Kardar-Parisi-Zhang; CBL, convective boundary layer.}

\author[1]{Ziwei Li}
\author[1]{Paul A. O'Gorman}
\author[2]{Daniel H. Rothman}

\affil[1]{Department of Earth, Atmospheric and Planetary Sciences, Massachusetts Institute of Technology, 
Cambridge, MA, 02139, United States}
\affil[2]{Lorenz Center, Department of Earth, Atmospheric and Planetary Sciences, Massachusetts Institute of Technology, 
Cambridge, MA, 02139, United States}

\corraddress{Ziwei Li, Department of Earth, Atmospheric and Planetary Sciences, Massachusetts Institute of Technology, 
Cambridge, MA, 02139, United States}
\corremail{ziweili@mit.edu}

\fundinginfo{
US National Science Foundation, Grant Number: AGS 1552195 and AGS 1749986; 
mTerra Catalyst Fund}

\runningauthor{Li et al.}

\begin{document}

\maketitle

\begin{abstract}
Tropical precipitation clusters exhibit power-law frequency
distributions in area and volume (integrated precipitation), implying a
lack of characteristic scale in tropical convective organization.  
However, it remains unknown what gives rise to the power laws and
how the power-law exponents for area and volume are related to one another.
Here, we explore the perspective that precipitation clusters are islands 
above a convective threshold on a rough column-water-vapor (CWV) topography.  
This perspective is supported by the agreement between the precipitation clusters and CWV islands 
in their frequency distributions as well as fractal dimensions.  
Power laws exist for CWV islands at different thresholds through
the CWV topography, suggesting that the existence of power-laws is not
specifically related to local precipitation dynamics, but is rather a general feature of CWV islands. 
Furthermore, the frequency distributions and fractal dimensions of the clusters can be
reproduced when the CWV field is modeled to be self-affine with a roughness exponent of 0.3. 
Self-affine scaling theory relates the statistics of precipitation clusters to the roughness exponent; it also
relates the power-law slopes for area and volume without involving the roughness exponent.  
Thus, the perspective of precipitation clusters as CWV islands provides a useful framework to consider
many statistical properties of the precipitation clusters, particularly given that CWV is well-observed over a wide range of length scales in the tropics.  
However, the statistics of CWV islands at the convective threshold imply a
smaller roughness than is inferred from the power spectrum of the bulk CWV field, 
and further work is needed to understand the scaling of the CWV field.

% Please include a maximum of seven keywords
\end{abstract}

\linenumbers

\section{Introduction}

%Tropical precipitation accounts for a substantial fraction of Earth's total precipitation. 
%The associated latent heat release and cloudiness play an important role in Earth's energy budget \citep{Tian2002} while intense tropical precipitation events are expected to increase in frequency with climate change \citep{OGorman2012}.

Tropical convection and associated precipitation are organized in clusters of 
spatial scales from 10 to 1000 km \citep[e.g.,][]{Mapes1993, Quinn2017a}. 
Understanding this organization is important because of the societal impacts of the spatial patterns of tropical precipitation, 
the influence of convective organization on the  large-scale properties of the tropical atmosphere \citep{Tobin2012}, 
and the need to represent organization in convective parameterizations in global climate models \citep{Mapes2011}.
Furthermore, both mean and extreme tropical precipitation are expected to 
experience substantial change with global warming \citep{OGorman2012, Duffy2020}, 
in which convective organization could play an important role \citep{Rossow2013, Tan2015}.

Many studies have investigated the cause of the spatial clumping of convection in the idealized setting of radiative convective equilibrium
\citep{Bretherton2005, Muller2012, Craig2013, Emanuel2014, Wing2014, Wing2015}. 
This behavior is termed convective self-aggregation, and the physical processes that lead to self-aggregation in radiative
convective equilibrium are also thought to be active in the tropical atmosphere
\citep{Holloway2017,Beucler2019}.
The paradigm of self-aggregation shows that convection can organize even in the 
absence of surface temperature gradients and background shear, but it does not
by itself explain the spatiotemporal characteristics of convection and
precipitation found in the tropics.  In particular, numerous studies
have found power-law distributions of precipitation clusters in observations
\citep{Lovejoy1985,Peters2010, Peters2012, Quinn2017a, Teo2017}, GCM
simulations \citep{Quinn2017b}, and high-resolution simulations with explicit
convection \citep{OGorman2021}. 
Power-law distributions feature a probability density distribution of the form  
\begin{equation}
\Pr(x)\propto x^{-\tau}, 
\label{eq:power_law}
\end{equation}
where $\Pr(x)$ is the probability density of $x$. 
We refer to $\tau$ as the power-law exponent, and $\tau$ is positive in all cases throughout this paper.
Eq.\;\eqref{eq:power_law} is linear in the log-log space, and 
it is the only scale-invariant distribution in the sense that the distribution 
does not have a characteristic length scale \citep{Turcotte1992}.  % First chapter, page 2 in Turcotte, 1992
Therefore, the presence of power-law distributions suggests that precipitation is scale-free.  

Scale-invariance has been associated with critical systems in statistical
physics, including equilibrium critical phenomena \citep{Pathria2011} and
self-organized criticality (SOC) in forced-dissipative systems
\citep{Pruessner2012}.  These critical systems are characterized by divergence in 
correlation length and power-law distributions of quantities such as the cluster
area of positive magnetization in the Ising model \citep{Toral1987} and the
distribution of avalanche size and duration in SOC sandpile models
\citep{Bak1987, Pruessner2012}.  Therefore, when power-law distributions were
found in temporally- and spatially-connected precipitation clusters,
hypotheses were made that atmospheric precipitation is an instance of SOC
\citep{Peters2006,Neelin2008,Teo2017, Haerter2019}.  
The power-law distribution of temporal cluster volume (precipitation integrated
in time) would then correspond to the power-law distribution of avalanche size in SOC.
Apart from the power laws, the analogy extends further as the atmosphere slowly
builds up its water vapor via evaporation and has a sudden avalanche-like onset
of precipitation once the column water vapor reaches a critical value
\citep{Peters2006, Neelin2008}.  However, it remains unclear whether common SOC
models \citep[e.g.,][p. 82]{Pruessner2012} can explain the observed power-law
exponents of precipitation clusters.  

%Another body of work has sought to build simple stochastic models that capture the main features of precipitation statistics in the tropics. The power laws associated with temporal precipitation clusters have been explained using simple stochastic models with a threshold \citep{Stechmann2011, Stechmann2014}, and the sudden onset of precipitation as a function of column-integrated water vapor (CWV) has been related to a two-variable stochastic model assuming independent Gaussian distributions of water vapor in the boundary layer and in the free troposphere \citep{Muller2009}.  Similarly, the spatial clustering of precipitation was simulated by stochastic reaction-diffusion equations \citep{Hottovy2015, Ahmed2019}.

In this paper, we focus on spatial precipitation clusters
that are defined as groups of precipitating grid points connected in the horizontal.
Cluster \textit{area} is defined as the horizontal area of the cluster, 
and cluster \textit{volume} is defined as the spatially integrated precipitation rate over the cluster
following \cite{Quinn2017a}, although they converted volume to an equivalent "power" associated with latent heating.
Frequency distributions of precipitation clusters exhibit power laws with exponents in the range of 2.0 to 1.7 
for area \citep{Lovejoy1985,Peters2009,Peters2012,Quinn2017a,Teo2017} and 1.7 to 1.5  for volume \citep{Quinn2017a, Teo2017}.
The spatial clustering of precipitation has been simulated by stochastic reaction-diffusion equations \citep{Hottovy2015, Ahmed2019}. 
The stochastic model of \cite{Ahmed2019} includes representations of precipitation and lateral moisture transport, 
and it produces frequency-distribution exponents of 1.6 for area and 1.5 for volume, which are close to the observed exponents. 
To explain the exponents, \cite{Ahmed2019} used a stochastic branching process which gives 
the same exponent of 1.5  for both area and volume, although a direct connection 
between the branching process and precipitation processes was not provided. 

From a different perspective, \citet{Pelletier1997} proposed that the frequency distribution of tropical cumulus cloud area 
could be understood through the statistical properties of the convective boundary layer (CBL) height. 
The CBL height field was taken to be a self-affine surface, and 
clouds were assumed to form wherever the CBL height exceeds a certain threshold. 
The self-affine scaling theory of \citet{Kondev1995} was then used to relate the 
area-distribution of clouds to the roughness of the CBL height field.  
\citet{Pelletier1997} further hypothesized that the roughness exponent of the CBL 
height field has a value of 0.4 because of Kardar-Parisi-Zhang (KPZ) dynamics \citep{Kardar1986}. 
This roughness could also be connected to the fractal dimension of clouds, 
which was previously found to be 1.35 \citep{Lovejoy1982}, 
although clouds a have slightly different dimension for length scales below 1 km \citep{Benner1998}.  
The same fractal dimension of cumulus clouds has alternatively been related to 
three-dimensional turbulence \citep{Siebesma2000} and gradient percolation theory \citep{Peters2009}.

We take a somewhat similar approach to \cite{Pelletier1997}
in that we seek to understand clusters based on a threshold through a rough surface. 
However, we consider precipitation clusters rather than cumulus clouds, 
and we relate the clusters to the field of column-integrated water vapor (CWV) rather than CBL height. 
CWV has units of mm and represents the height of liquid water if all water vapor in the column is condensed onto the surface. 
Using CWV has the advantage that it is readily observed over a wide range of length scales.  
Furthermore, precipitation undergoes a rapid pickup once the column-integrated water vapor 
(CWV) exceeds a critical value as seen in
observations \citep{Peters2006, Raymond2009, Neelin2009, Ahmed2015} and simulations
\citep{Bretherton2005, Sahany2012, Yano2012, Posselt2012}. 
This property has been used in the stochastic models of \cite{Hottovy2015} and \cite{Ahmed2019}. 
The sharp pickup occurs because of the conditional instability of moist convection, 
as moist convection tends to occur with abundant low-level moisture through moist air parcels rising from near the
surface and abundant mid-level moisture due to the effects of entrainment \citep{Holloway2009,
Muller2009}. We expect that the critical CWV to have weak variations in the horizontal due to
the weak horizontal temperature gradients in the tropical free troposphere. 

\begin{figure}[h!]
\centering
\noindent\includegraphics[width=1.0\linewidth]{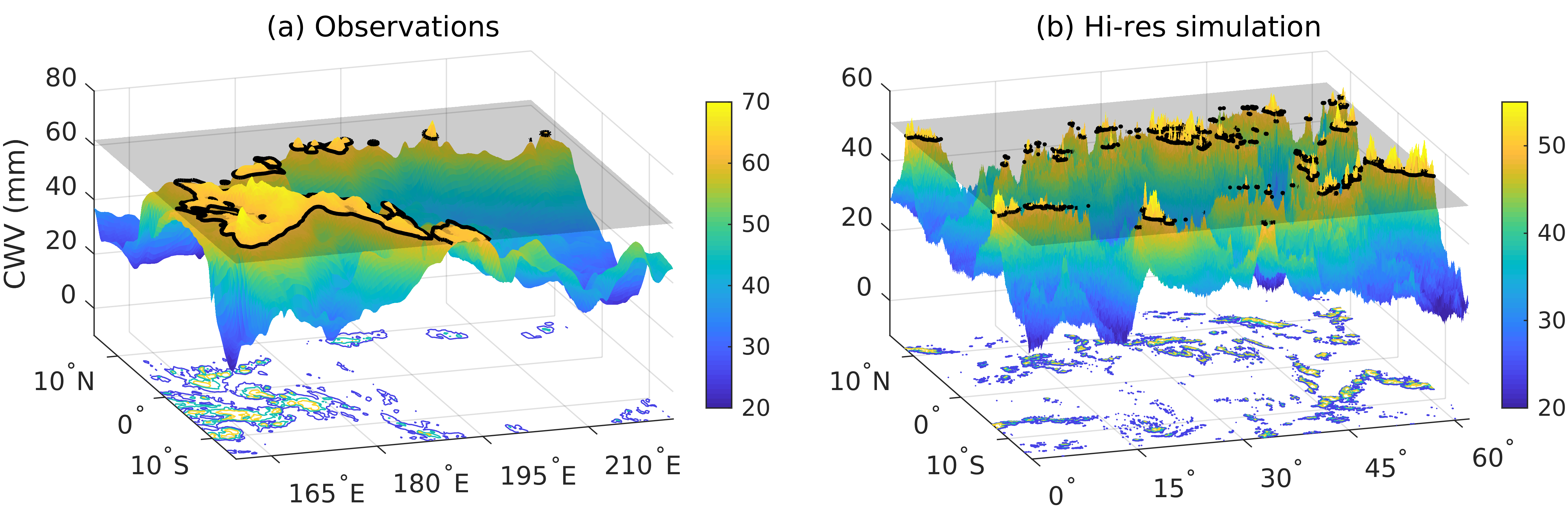}
\caption{Examples of precipitation clusters as islands on a rough CWV topography using (a) observations and (b) a high-resolution simulation. 
The observations are taken from TRMM-3B42 for precipitation and ERA5 reanalysis for CWV. 
In both panels, the colored surfaces aloft show 3-hourly averaged CWV, 
the contours at the bottom show accumulated precipitation in the same 3-hour period, 
and the blue, green, and yellow contours correspond to precipitation rates of 10, 50, and 90 mm day$^{-1}$. 
The transparent planes represent convective CWV thresholds of 62 mm for (a) and 51 mm for (b), 
and the thick black contours highlight CWV island perimeters. }
\label{fig:topography}
\end{figure}

Here, we regard precipitation clusters as manifestations of CWV \textit{islands} 
above a fixed \textit{threshold} on a rough CWV topography (Fig.\;\ref{fig:topography}). 
The fixed threshold is the convective threshold of CWV above which precipitation rapidly increases. 
The power-law frequency distribution of precipitation cluster area is then akin to 
the Kor\v{c}ak's law, which describes a power-law distribution of island area above sea level on Earth's relief \citep{Mandelbrot1982, Imre2015}.  
We further assume that precipitation is linear in the excess of CWV above the threshold, 
such that the volume of a precipitation cluster corresponds to the volume of the CWV island above the threshold. 
Consistent with prior studies which show that power-law distributions for islands on a rough topography are a generic result \citep[e.g., ][]{Olami1996}, 
we find that the existence of power-law CWV island distributions is not dependent on 
the choice of the threshold and is not expected to be tied to the specific dynamics in precipitating regions. 

From a combination of observations and simulations, we show that tropical
precipitation clusters are closely connected to CWV islands in
area and volume frequency distributions and also in fractal dimensions.
We further generate idealized self-affine surfaces and 
show that the thresholded islands on these surfaces correspond well with the statistics of CWV islands. 
Assuming that the CWV field is self-affine allows us to apply the self-affine scaling
theories for contour loops of \citet{Kondev1995} and \citet{Kondev2000} to predict the
exponents of cluster area and volume distributions.  We show
how these exponents and the cluster fractal dimensions 
can be related to the roughness exponent of the CWV field and to each
other. While the self-affine scaling theory a useful framework, its
quantitative predictions are not very accurate in some cases because the
CWV is not exactly self-affine, and because the scaling theory
is derived for all contour loops at all thresholds, not for contour loops at a single threshold. 

This paper is organized as follows.  We describe the frequency distributions of
precipitation clusters and their fractal dimensions in observations and simulations in section \ref{sec:power_law}. 
We then demonstrate the similarity between the statistics of precipitation clusters and thresholded CWV islands in section \ref{sec:clusters}.  
We further make the idealization that CWV is self-affine and apply the self-affine scaling theory 
to give expressions for the power-law distribution exponents and fractal dimensions of CWV islands in section \ref{sec:theory}. 
Lastly, we give our conclusions in section \ref{sec:conclusion}.

\section{Distributions and dimensions of precipitation clusters in different datasets}\label{sec:power_law}

We analyze precipitation and CWV statistics in observations, a high-resolution 
simulation with explicit convection (hereby hi-res), and a GCM simulation.  
For observations, we use precipitation from TRMM-3B42
\citep{Huffman2007} and CWV from the ERA5 reanalysis \citep{Hersbach2020}, both
of which are on a 0.25$^\circ$ by 0.25$^\circ$ grid. We refer to
ERA5 CWV as observations for simplicity even though it is from a reanalysis dataset. 
For the hi-res simulation, we use the system for atmospheric modeling (SAM)
\citep{Khairoutdinov2003}, configured as a semi-global
aquaplanet on an extended equatorial beta plane with a hemispherically- and zonally-symmetric 
sea surface temperature distribution.  The domain spans from
78$^\circ$S-78$^\circ$N in latitude and 62$^\circ$ in longitude at the equator.
The horizontal grid spacing is 12 km, and hypo-hydrostatic rescaling 
\citep{Kuang2005, Garner2007, Fedorov2018} is applied to reduce the horizontal
scale difference between convection and large-scale dynamics.  
See \cite{Yuval2020} and \cite{OGorman2021} for more details of hi-res. 
For the GCM simulation, ensemble number 1 in the CESM large ensemble dataset \citep{Kay2015} 
is used as a representative coupled atmosphere-ocean GCM simulation,
which has a grid spacing of 1.25$^\circ$ in longitude and 0.94$^\circ$ in latitude. 
The observations and GCM datasets span a period from 01/01/2002 to 12/31/2005
which is the longest overlap of the two datasets, 
and the hi-res simulation has a simulation length of 1200 days. 
The precipitation rate and the CWV field are 3-hourly averaged for observations and hi-res. 
For the GCM simulation, the precipitation rate is 6-hourly averaged, 
and the CWV field is calculated using a mass-integral of its 6-hourly instantaneous specific humidity output. 
All results presented are based on a region of 15$^\circ$S-15$^\circ$S, 160$^\circ$E-222$^\circ$E 
in the central tropical Pacific for observations and GCM, 
and 15$^\circ$S-15$^\circ$S with all available longitudes for hi-res. 

We define precipitation clusters as groups of precipitating grid points that are connected
via nearest-neighbor bonds, where there are four nearest-neighbors to each grid point. 
Precipitating grid points are grid points where the precipitation rate exceeds 0.7 mm h$^{-1}$. 
This precipitation threshold is chosen to be consistent with prior works such as \cite{Quinn2017a}. 
Using a different threshold between 0.1 and 2.5 mm h$^{-1}$ does not noticeably change the shape of the cluster distributions.
Consistent with Fig.\;2d in \citet{Otsuka2017}, the cluster area distribution becomes lognormal-like 
when a much higher threshold of 20 mm h$^{-1}$ is used,
which might explain why some previous studies found that cloud clusters follow a lognormal distribution \citep[e.g., ][]{Mapes1993}. 

Following Eq.\;\eqref{eq:power_law}, we denote the power-law exponents for cluster area and volume 
distributions as $\alpha$ and $\beta$, respectively, where $\alpha$ and $\beta$ are positive when the log-log slope is negative. 
The meanings of all symbols used in the paper are summarized in Table \ref{tab:exponent_meaning}.  
To estimate $\alpha$ and $\beta$, we sort cluster area and volume into 25 bins and apply linear regression in the log-log space. 
We use logarithmic binning because it reduces the noise in the tail of the distribution \citep{Bauke2007}. 
The widths of the bins are rounded to the nearest multiples of the smallest area or volume, following \cite{Quinn2017a}. 
Each distribution's regression range is chosen based on the apparent extent of the power-law range. 
We report the error of each exponent in parenthesis after its estimated value. 
To obtain the error, we allow the starting bin to move upward by one bin, remain the same, 
or move downward by one bin, giving 3 choices of the starting point. 
The same applies to the end  bin, and together these choices yield 9 exponent values. 
We regard the largest absolute deviation out of the 9 values from the estimated value as the
measurement error of each exponent. This error dominates over the traditional standard error of regression slope,
and we use it to represent the uncertainty in the measured exponents. 
The regression ranges of cluster volume distributions are approximately matched to those of 
the cluster area distributions in the sense that
they cover the same fractional distrance between the smallest and largest bins of the distribution in the log space.\footnote{The smallest and largest bins of the distribution are determined by the minimum and maximum of the clusters, and the start and end point of the regression has the same linear location in the log space relative to the largest and smallest bins. In practice, we use the same set of consecutive bins out of all 25 bins as the regression range for cluster area and cluster power distributions. }

\begin{figure}[h!]
\centering
\noindent\includegraphics[width=0.85\linewidth]{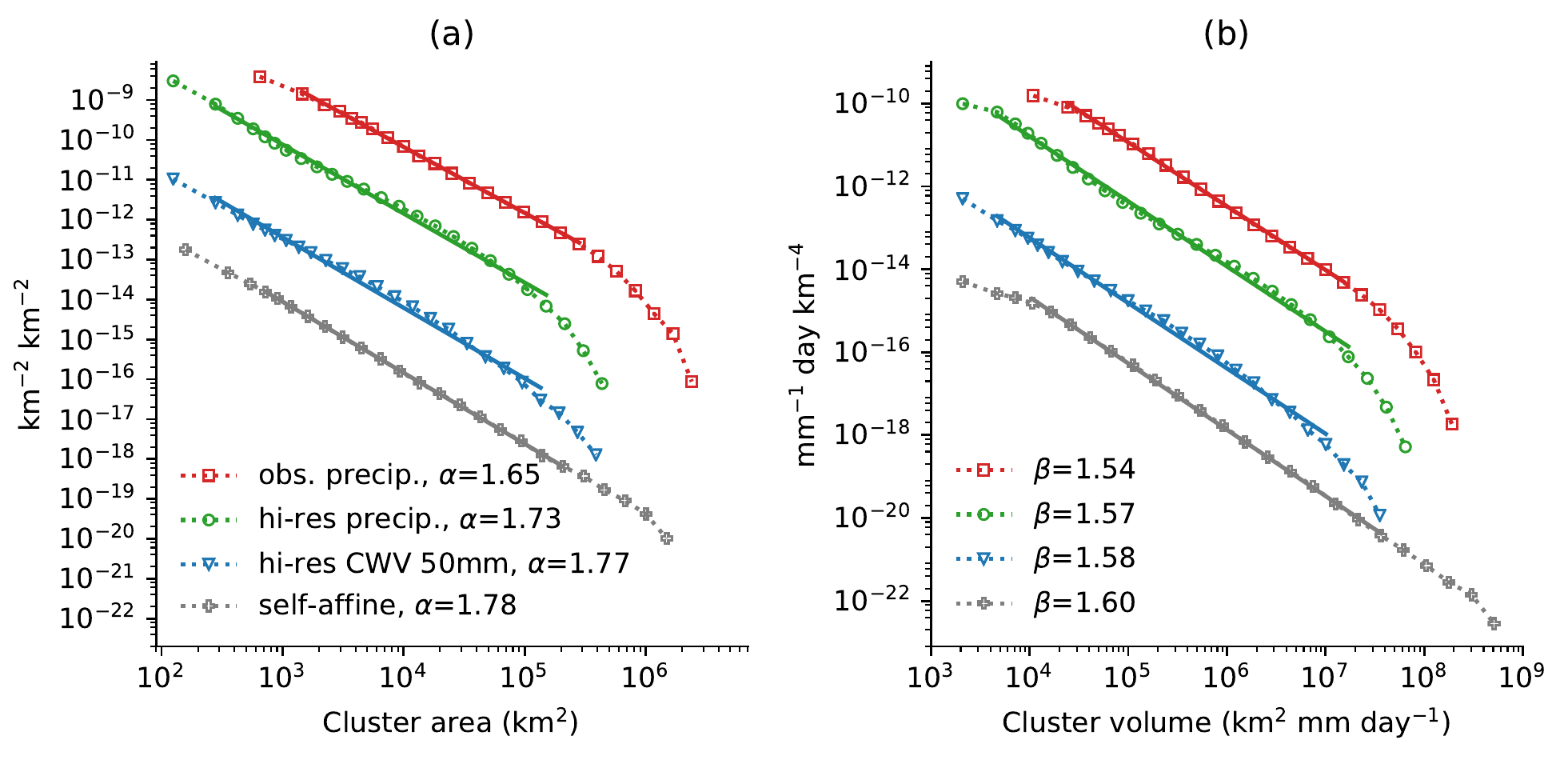}
\caption{The frequency distributions of (a) cluster area and (b) cluster volume. 
Different colors and markers correspond to observed precipitation (red squares), hi-res precipitation (green circles), 
hi-res CWV islands at 51 mm (blue triangles), 
and islands on the self-affine surfaces with $H$ = 0.3 at a threshold of 51 mm (gray crosses). 
The distributions are normalized such that the integral is the time-mean of the number of clusters per unit area of the domain, 
and they are consecutively shifted downwards by a factor of 50 starting from the observed precipitation for clarity. 
In (b), the volume of hi-res CWV islands and self-affine islands are converted to precipitation 
by Eq.\;\eqref{eq:precipitation} to plot island volume and precipitation volume on the same graph. 
The statistics are based on the region of 160$^\circ$E-222$^\circ$E and 15$^\circ$S-15$^\circ$N for observations, 
the same latitudinal band for hi-res, and the whole domain for self-affine. 
The solid lines are linear regressions in the log-log space, and their extents correspond to the regression ranges. 
}
\label{fig:distributions}
\end{figure}

Consistent with previous studies, we find power-law frequency distributions
with exponential upper cutoffs for precipitation cluster area and volume in
observations and hi-res (Fig.\;\ref{fig:distributions}).  
The cluster area exponent $\alpha$ is 1.65 (0.04)
for observations with a similar value of 1.73 (0.05) for hi-res. 
The parentheses after each exponent indicates the regression error as described above. 
The values and errors of all exponents for different datasets in this paper are summarized in table \ref{tab:exponents}. 
The cluster volume exponent $\beta$ is lower at
1.54 (0.04) for observations with a similar value of 1.57 (0.04) for hi-res.  
These values for $\alpha$ and $\beta$ are similar to values in previous studies that
also analyzed TRMM-3B42 \citep{Quinn2017a, Teo2017}.  
We regard the hi-res simulation as having an
idealized yet faithful representation of tropical precipitation 
\citep{OGorman2021}, and we will focus on hi-res from here on.  
The cluster distributions for GCM are different, and they are discussed in 
Appendix \ref{sec:GCM_power_spectrum}. 

\begin{table}[!h]
\caption{Precipitation cluster area ($\alpha$) and cluster volume ($\beta$) exponents, distribution scaling relation (Eq.\;\ref{eq:scaling_relation}), perimeter dimension ($D_l$), volume dimension ($D_V$), and dimension scaling relation (Eq.\;\ref{eq:dimension_relation}) for different datasets. The results for hi-res CWV and the self-affine surface are calculated for islands cut by a threshold at 51 mm, about 2.0$\sigma$ above the mean. }
\centering
\begin{tabular}{lcccccc}
\headrow
			&\thead{Obs. precip.} &\thead{Hi-res precip.} &\thead{Hi-res CWV} &\thead{Self-affine, H = 0.3}&\thead{Theory, H = 0.3}\\
$\alpha$              &1.65 (0.04)  &1.73 (0.05)  &1.77 (0.05)  &1.78 (0.01)  &1.85 (Eq.\;\ref{eq:area_distribution}) \\
$\beta$               &1.54 (0.04)  &1.57 (0.04)  &1.58 (0.04)  &1.60 (0.03)  &1.74 (Eq.\;\ref{eq:volume_distribution}) \\
$\alpha+2/\beta$      &2.95 (0.07)  &3.00 (0.08)  &3.04 (0.09)  &3.03 (0.03)  &3 (Eq.\;\ref{eq:scaling_relation}) \\
$D_l$                 &1.37 (0.02)  &1.41 (0.02)  &1.35 (0.02)  &1.39 (0.01)  &1.35 (Eq.\;\ref{eq:perimeter_dimension}) \\
$D_V$                 &2.32 (0.02)  &2.33 (0.04)  &2.32 (0.03)  &2.41 (0.01)  &2.3 (Eq.\;\ref{eq:volume_dimension}) \\
$2D_l+D_V$            &5.07 (0.06)  &5.14 (0.08)  &5.02 (0.07)  &5.18 (0.02)  &5 (Eq.\;\ref{eq:dimension_relation})  \\
\hline
\end{tabular}
\label{tab:exponents}
\end{table}

\iffalse
% Simplified table for presentation
\begin{table}[!h]
\centering
%\ra{1.}
\begin{tabular}{lcccccc}
\toprule[0.1em]
				&$\alpha$	&$\beta	$	&$\alpha+\frac{2}{\beta}$\\
\midrule[0.07em]
observed preciptation              		&1.65  	&1.54  	&2.95    \\
hi-res precipitation              		&1.68	 	&1.56 		&2.96    \\
hi-res CWV islands   		&1.71  	&1.52 		&3.01    \\
\bottomrule[0.1em]
\end{tabular}
\end{table}
\fi

\begin{figure}[h!]
\centering
\noindent\includegraphics[width=1.0\linewidth]{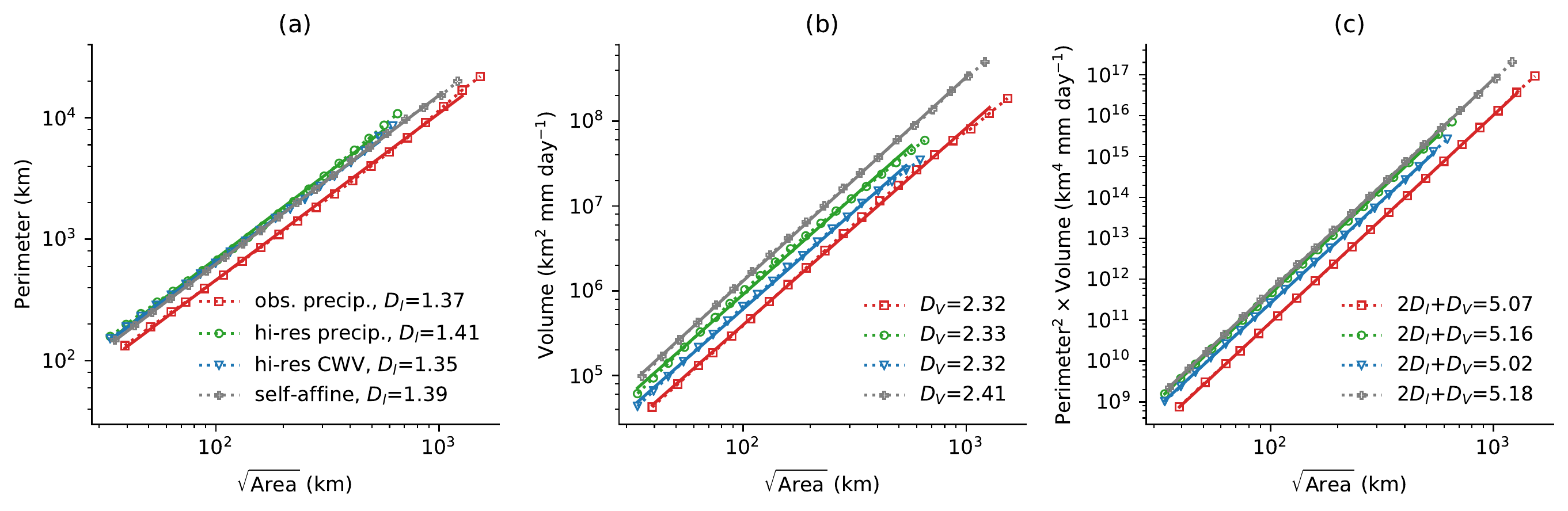}
\caption{(a) Perimeter, (b) volume and (c) perimeter squared multiplied by volume 
as functions of the square root of area for observed precipitation clusters (red squares), 
hi-res precipitation clusters (green circles), hi-res CWV islands at 51 mm (blue trangles), 
and islands on a self-affine surface with $H$ = 0.3 at 51 mm (gray crosses). 
Solid lines show linear regressions in the log-log space with the estimated slopes in the legends. 
In (b) and (c), the volume of hi-res CWV islands and self-affine islands are converted to precipitation by Eq.\;\eqref{eq:precipitation}. }
\label{fig:cluster_dimensions}
\end{figure}

We use the area-perimeter scaling to estimate the fractal perimeter dimension of precipitation clusters (Fig.~\ref{fig:cluster_dimensions}a). 
This self-similar scaling was first adopted to study fractal cloud dimensions by \cite{Lovejoy1982}. 
For a set of two-dimensional self-similar fractal objects, their perimeter length is related to area and radius by 
\begin{equation}
l \propto A^{D_l/2} \propto R^{D_l}, 
\end{equation}
where $l$ is perimeter length, $A$ is area, $D_l$ is the perimeter dimension, and 
$R$ is the radius which can be thought of as the edge of the smallest square that can cover the object. 
The perimeters are traced out using \textit{find\_contours()} in the \textit{scikit-image} library which 
implements a two-dimensional version of the marching cubes algorithm \citep{Lorensen1987}. 
$D_l$ is determined by binning $\sqrt{A}$ in the log space, taking the average of $l$ in each bin, 
and regressing the $l$ averages against $\sqrt{A}$ in the log-log space. 
The regression ranges used are indicated by the extents of the solid lines in Fig.\;\ref{fig:cluster_dimensions}.
The uncertainty in the regression slopes are estimated in the same way as for the frequency distributions by 
varying the start and end point upwards or downwards by one bin and finding the maximum deviation. 
The $D_l$ of precipitation clusters is 1.37 (0.02) for observations and has a similar value of 1.41 (0.02) for hi-res. 
These values are also broadly consistent with previous findings that the fractal dimension of cloud perimeter
is 1.35 for radii from 1 to 1000 km \citep{Lovejoy1982}.

We also investigate the scaling of cluster volume with area (Fig.\;\ref{fig:cluster_dimensions}b).
We introduce a volume fractal dimension, $D_V$, such that
\begin{equation}
    V\propto A^{D_V/2} \propto R^{D_V}. 
    \label{eq:D_V_definition}
\end{equation}
The precipitation clusters in observations and hi-res have similar $D_V$ values of 2.32 (0.02) and 2.33 (0.04), 
respectively, with the dimensions and errors estimated using the same approach as for $D_l$, $\alpha$, and $\beta$.

\section{Precipitation and thresholded CWV clusters}\label{sec:clusters}

\begin{figure}[h!]
\centering
\noindent\includegraphics[width=0.35\linewidth]{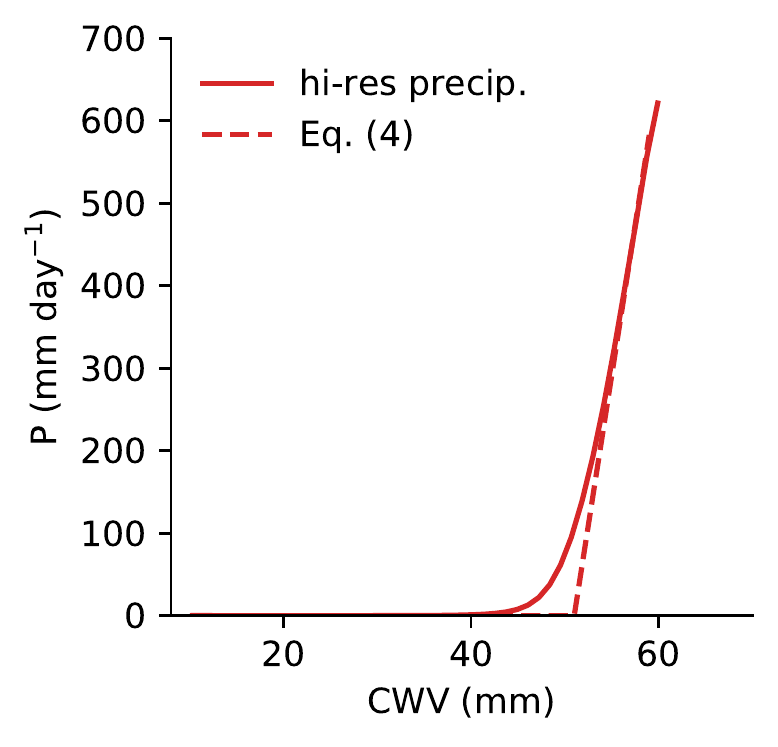}
\caption{Tropical precipitation binned by CWV in the hi-res simulation (solid) 
and the estimate of Eq.\;\eqref{eq:precipitation} with $C$ = 72.8 day$^{-1}$ and $Q_c$ = 51 mm (dashed). 
Bins which have less than a millionth of the total number of data instances are not shown. }
\label{fig:precip_pickup}
\end{figure}

To better understand the statistical properties
of precipitation clusters, we envision them as islands above a convective
threshold on a rough CWV topography. 
Denoting CWV as $Q$, we define a CWV \textit{convective threshold}, $Q_c$, which quantifies convective inhibition. 
We assume that the precipitation rate is zero when CWV is below $Q_c$, and the precipitation rate 
scales linearly with the excess of CWV when CWV is above $Q_c$: 
\begin{equation}
P(\mathbf{r}) = 
\left\{ \begin{aligned} 
  & C(Q -  Q_c)~~\text{when}~Q>Q_c, \\
  & 0 ~~~~~~~~~~~~~~~~~~~\text{otherwise}.
 \end{aligned} \right.
\label{eq:precipitation}
\end{equation}
$P(\mathbf{r})$ is the 3-hourly precipitation rate at location $\mathbf{r}$, and $C$ is a proportionality factor. 
The value of $C$ does not affect the analytical results of the power-law exponents
or fractal dimensions in later sections.
Eq.\;\eqref{eq:precipitation} can be thought of as a first-order parameterization that captures the onset of precipitation once CWV exceeds a threshold. 
Fig.\;\ref{fig:precip_pickup} shows the mean precipitation rate conditioned on CWV, 
in which CWV values are binned with constant intervals in the linear space, 
and the precipitation rate is averaged in each bin. 
We find that Eq.\;\eqref{eq:precipitation} works well for 
the hi-res simulation as shown in Fig.\;\ref{fig:precip_pickup}, while noticing the fact that 
the exact functional form relating precipitation to CWV differs to some
extent across different observational and modeling studies 
\citep{Neelin2009, Sahany2012, Yano2012, Posselt2012, Ahmed2015}

The threshold $Q_c$ cuts through the CWV field and gives a
collection of distinct islands above the threshold (Fig.\;\ref{fig:topography}). 
With Eq.\;\eqref{eq:precipitation}, each CWV island has an associated hypothetical precipitation cluster. 
We define the volume of the CWV island as the volume of the hypothetical precipitation cluster,
which is the spatial integral of $C(Q -  Q_c)$ within the island. 
The island's projected area is the area of the hypothetical precipitation cluster.  

We choose a CWV threshold of 51 mm for hi-res throughout the paper because this
is the integer threshold that gives CWV islands with both $\alpha$ and $\beta$
closest to the precipitation clusters as discussed below.  This threshold is
also close to the value of 48 mm that gives the best match for the mean area
fraction of precipitation in the hi-res simulation.  We determine the
proportionality factor, $C$, by regressing the linear part of Eq.\;\eqref{eq:precipitation}
against the bin-averaged precipitation rates in Fig.\;\ref{fig:precip_pickup}. 
For hi-res, the CWV threshold of 51 mm gives $C$ = 72.8 day$^{-1}$, 
meaning that 1 mm in CWV above the threshold corresponds to 72.8 mm
day$^{-1}$ in precipitation.  A higher threshold of 62 mm is chosen for the case of
observations, but it is only used for illustration in
Fig.\;\ref{fig:topography}(a).  We don't match the distribution of CWV islands
to the distribution of precipitation clusters in observations because the
moisture field from the ERA5 reanalysis is smooth at small
length scales, making its CWV island distributions not power-law-like at high CWV thresholds. 
%\footnote{The deviation of the CWV island distribution in ERA5 reanalysis from a power law could also be related to variations in sea surface temperature. Large surface temperature corresponds to large saturation water content in the column, thus making a fixed CWV threshold not an ideal approximation.}. % largest variation in SAM is 2K, whereas it's easily 5K in ERA5 reanalysis for the selected region
The threshold for hi-res is lower than for observations because the average sea
surface temperature in hi-res is lower than that in observations in the selected central tropical Pacific domain. 

To support the notion that precipitation clusters are manifestations of
thresholded CWV islands, we first directly compare the pattern of thresholded CWV islands
to that of precipitation clusters in observations and hi-res in Fig.\;\ref{fig:topography}. 
CWV islands in both observations and hi-res have very similar shapes to precipitation clusters. 
There is a dominant CWV island accompanied by multiple smaller islands in observations (Fig.\;\ref{fig:topography} a), 
whereas multiple medium-area islands prevail in hi-res (Fig.\;\ref{fig:topography} b); the same pattern also goes for precipitation clusters. 
This difference in CWV island (and precipitation cluster) configuration is due to the tropical Pacific warm pool being located on the
western side in the domain of observations while the sea surface temperature is zonally uniform for hi-res. 

Hi-res CWV islands also have power-law distributions in area and volume, 
and the power-law exponents are close to those of the precipitation clusters (Fig.\;\ref{fig:distributions}).  
To generate CWV island distributions, we randomly
sample 500 snapshots of 3-hourly averaged CWV field of hi-res. The hi-res simulation is used instead of
observations or GCM because hi-res has the highest resolution and doesn't show
evidence of smoothing in the CWV field at small length scales.  
We set the CWV island volumes that are smaller than the minimum precipitation cluster volume, 2419.2 km$^2$ mm day$^{-1}$, to 
2419.2 km$^2$ mm day$^{-1}$.\footnote{The minimum volume, 2419.2 km$^2$ mm day$^{-1}$, is equal to having a precipitation rate of 0.7 mm h$^{-1}$ at a single grid point of size 144 km$^2$. The 0.7 mm h$^{-1}$ rate is the minimum precipitation rate used to define precipitation in section \ref{sec:power_law}.}
Otherwise, the plotting of CWV island distributions is exactly the same as for precipitation clusters.  

For the CWV threshold of 51 mm in hi-res, the frequency distributions of area
and volume of the CWV islands are a good match to those of precipitation
clusters for the power law ranges (Fig.\;\ref{fig:distributions}). 
The measured power law exponents are
$\alpha$ = 1.77 (0.05) and $\beta$ = 1.58 (0.04) for the CWV islands as compared to
$\alpha$ = 1.73 (0.05) and $\beta$ = 1.57 (0.04) for the precipitation clusters.  
The areas of the largest precipitation clusters and the largest CWV islands at the
convective threshold of 51 mm are similar at about 4$\times$10$^5$ km$^2$.  The mean and
standard deviation of the CWV field are 35.3 mm and 8.0 mm, respectively, so
that 51 mm is roughly 2.0$\sigma$ above the mean value.  

The fractal dimensions of the hi-res CWV islands at 51 mm are also in good agreement 
with those of the hi-res precipitation clusters (Fig.\;\ref{fig:cluster_dimensions}a, b).  
The $D_l$ for CWV clusters at a threshold of 51 mm is 1.35 (0.02), slightly lower than the $D_l$ of 1.41 (0.02) for hi-res precipitation clusters. 
Similarly, the $D_V$ for CWV clusters at the 51 mm threshold is 2.32 (0.03), 
close to the $D_V$ of 2.33 (0.04) for precipitation clusters. 

\begin{figure}[h!]
\centering
\noindent\includegraphics[width=0.8\linewidth]{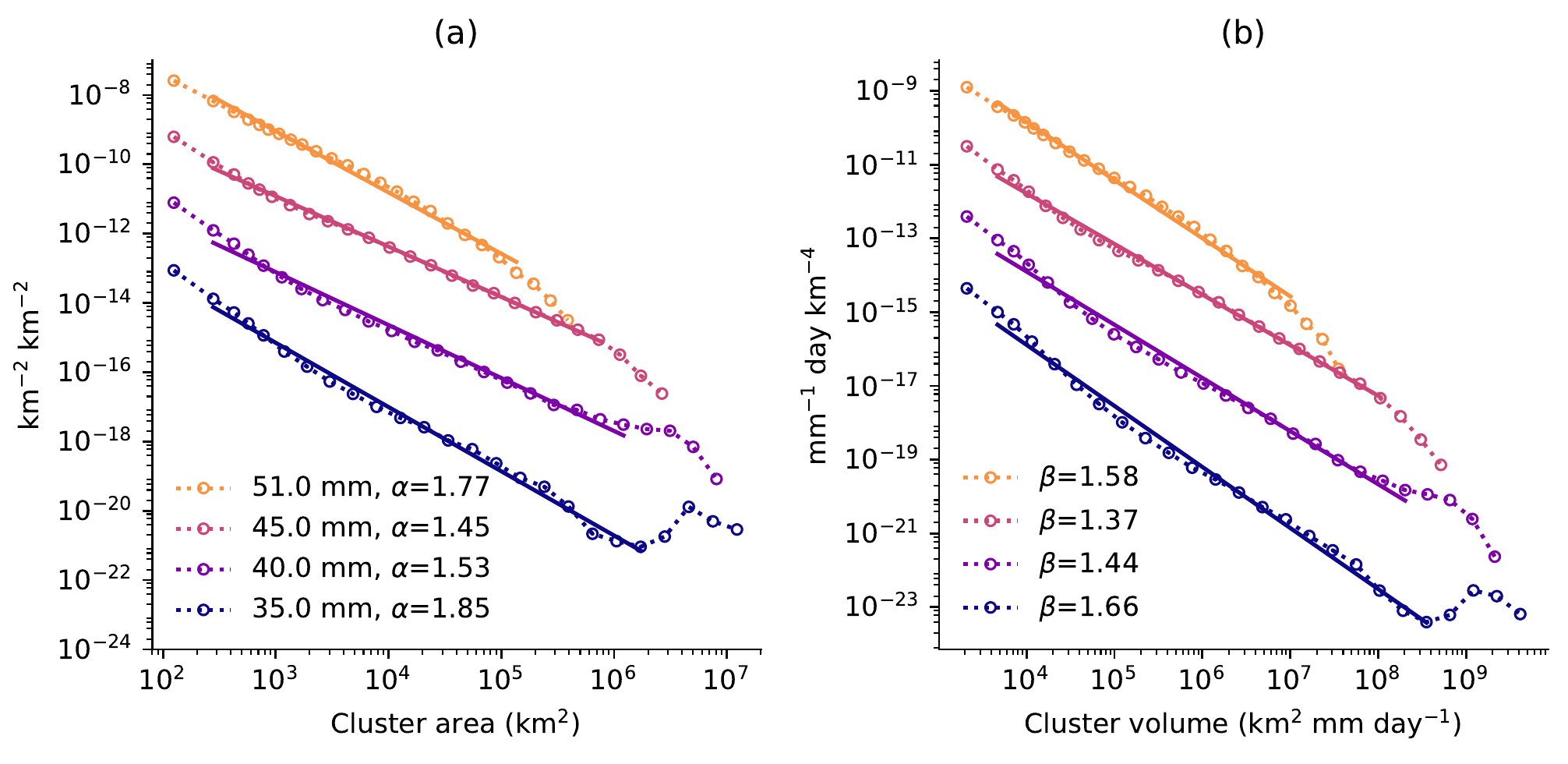}
\caption{Dotted lines with circles show the frequency distributions of (a) area and (b) volume of hi-res CWV islands at different thresholds 
of 51 mm, 45 mm, 40 mm, and 35 mm (from orange to navy, top to bottom). 
Solid lines show linear regressions in the log-log space, with the corresponding exponents in the legends. 
The regression ranges are indicated by the horizontal extent of the solid lines. 
The distributions are normalized as those in Fig.\;\ref{fig:distributions} and, starting from 51 mm, 
are consecutively shifted downwards by two decades for clarity. }
\label{fig:CWV_cluster_distributions}
\end{figure}

Interestingly, power-law distributions of CWV island area and volume exist for a wide range of CWV thresholds (Fig.\;\ref{fig:CWV_cluster_distributions}).
This is the case even for thresholds like 35 mm far below the convective threshold (51 mm), 
such that the precipitation rate over most of the island coverage is close to zero. 
The area and volume distributions for 35 mm have a much larger maximum area and volume
than the distributions for 51 mm because 35 mm cuts through a larger portion of the CWV topography as compared to 51 mm.  
Thus, the CWV island distributions at 35 mm has a local maximum at very large area and volume 
due to the presence of continents in the domain, and the maximum values in area and volume of
precipitation and CWV islands at the 51 mm convective threshold are smaller. 
Similar to the case of precipitation clusters
(Fig.\;\ref{fig:distributions}), $\alpha$ is larger than $\beta$ for
the distributions of CWV islands at different thresholds (Fig.\;\ref{fig:CWV_cluster_distributions}). 
$\alpha$ and $\beta$ are not
constant for different CWV thresholds. Rather, both exponents follow a similar trend
where they decrease and then increase as the CWV threshold is raised from
near the mean level of 35 mm to the convective threshold of 51 mm (Figs.\;\ref{fig:CWV_cluster_distributions} and 
%\ref{fig:CWV_exponents}). 
S1). 
The reasons for this variation are discussed in section \ref{sec:theory}. 

That the frequency distributions for the area and volume of CWV islands at the convective CWV threshold are very
similar to those of precipitation clusters and that their fractal dimensions
are also in good agreement suggest that tropical precipitation clusters are
manifestations of thresholded CWV islands and are in turn related to the CWV field. 
This allows us to use the geometric properties of the CWV field to
understand the existence of power laws and the relationships between $\alpha$,
$\beta$, $D_l$, and $D_V$.  The fact that power-law frequency distributions
exist for CWV islands at different thresholds also implies that the existence
of power laws does not depend on local precipitation dynamics such as gust
fronts or cold pools, but is more related to the scale-free nature of CWV dynamics
which occurs in both precipitating and non-precipitating regions in the
tropics.  On the other hand, precipitation dynamics may affect
the roughness of the CWV field and thus influence the power-law exponents of
the frequency distributions and fractal dimensions.  
In the next section, we use a self-affine scaling theory to obtain analytical
expressions that help explain the power-law exponents and fractal dimensions. 

\section{Applying self-affine scaling theory to the CWV topography}\label{sec:theory}

We seek theories that can predict the CWV island frequency distributions and fractal dimensions 
from the statistical properties of the CWV field, which in turn 
give predictions for the corresponding properties of precipitation clusters.  
The traditional percolation problem on a two-dimensional lattice does not explain the slope of the island area distribution because 
it only has a power-law area distribution at a single threshold, i.e., the percolation threshold, 
and the area distribution at this threshold has a power-law exponent of 187/91 $\approx$ 2 \citep[see table 2 in][]{Stauffer1994}, 
which is steeper than the area-distribution exponents ($\alpha$) in this paper. 

We observe that the perimeter and volume of CWV islands in hi-res
exhibit scaling relationships with area (Fig.\;\ref{fig:cluster_dimensions}a, b), 
and that the power spectrum of CWV approximately follow a power-law 
over a wide range of wavenumbers (Fig.\;\ref{fig:spectrum}).
These properties suggest that CWV may be modeled as a \textit{self-affine} surface \citep[e.g., ][]{Mandelbrot1985, Barabasi1995}. 
An isotropic self-affine surface, $h(\mathbf{r})$, satisfies
\begin{equation}
    h(\mathbf{r})\sim b^{-H}h(b\mathbf{r}), 
    \label{eq:self_affine}
\end{equation}
where $h(\mathbf{r})$ is surface height at location $\mathbf{r}$, $b$ is a rescaling factor, $H$ is the roughness exponent,  
and $\sim$ means statistical equivalence.
Eq.\;\eqref{eq:self_affine} states that the statistical properties of a subset of the surface (left side of the equation assuming $b > 1$) 
is the same as that of the surface itself (right side), subject to a rescaling of $b^{-H}$ in height. 
Typically, $H$ takes values between 0 and 1. For a fixed vertical width (standard deviation) 
at the largest horizontal scale of the system, the surface has less small-scale variations 
for larger $H$ \citep{Krim1993}. 

\subsection{Idealized self-affine surfaces}\label{sec:idealized_surface}

We first generate idealized self-affine surfaces to assess whether the islands on
these surfaces correspond well to the CWV islands.  The self-affine
surfaces are generated in a square domain with 512 points in each direction.  A
grid spacing of 13.5km is chosen such that the side of the square domain has
the same extent as the hi-res simulation in the zonal direction. 
The mean and standard deviation are chosen to match those of the hi-res CWV field.  The
self-affine surfaces are statistically isotropic with a power-law power spectrum 
$S(k) \propto k^{-\mu}$ where $k$ is the wavenumber and $\mu=2H+1$ \citep[Eq.\;7.48 in][]{Turcotte1992}.
We generate 500 surfaces, and for each surface, the phases of its Fourier components are randomly sampled in $[0, 2\pi)$
with a uniform distribution. The resulting surfaces also belong to Gaussian random
surfaces because the height field has a Gaussian distribution.

\begin{figure}[h!]
\centering
\noindent\includegraphics[width=0.8\linewidth]{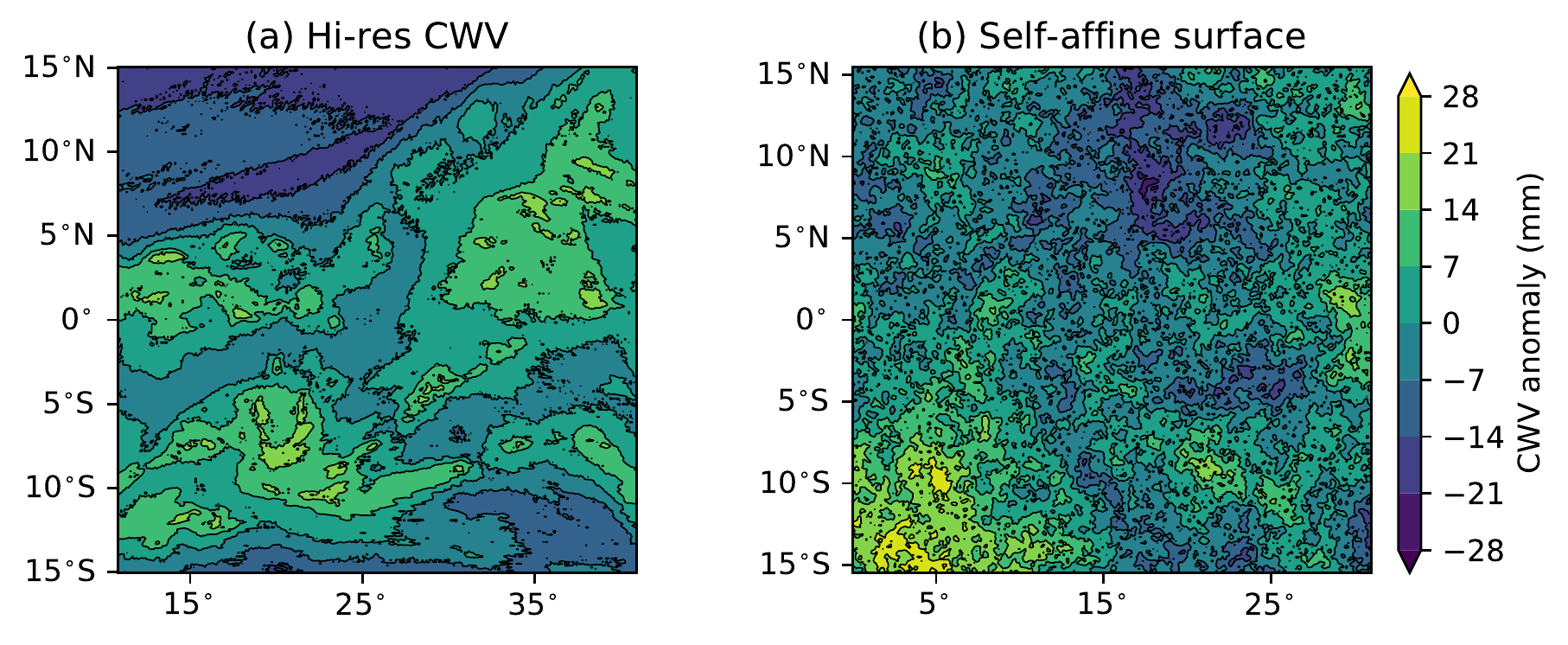}
\caption{Snapshots (shading) and the corresponding level sets (contours) of the anomalies of 
(a) hi-res CWV and (b) an idealized self-affine surface with $H$ = 0.3. 
Only a subset of the domain is shown in each case, and the spatial mean in each panel is removed for a better comparison. }
\label{fig:contours}
\end{figure}

We test a range of $H$ values and find that $H$ = 0.3 gives the best overall
agreement with the hi-res CWV field in terms of island frequency distributions
and island fractal dimensions at the convective threshold of 51 mm, or 2.0$\sigma$ above the mean. 
Interestingly, $H$ = 0.3 is close to the surface growth model of KPZ \citep{Kardar1986} 
which measures $H\simeq0.39$ in numerical simulations\footnote{There has not been an exact calculation of $H$ for KPZ in 2 dimensions. Numerical simulations in 2 dimensions seem to converge to $H\simeq0.39$ \citep{Pagnani2015}.} and was used to relate cumulus cloud distribution to convective boundary layer height \citep{Pelletier1997}. 
Fig.\;\ref{fig:contours}(b) shows an example of the generated self-affine surface.  
The area and volume frequency distributions of self-affine islands at the 51 mm threshold follow power laws (Fig.\;\ref{fig:distributions}). 
The exponents are $\alpha$ = 1.78 (0.01) and $\beta$ = 1.60 (0.03), respectively, which are
close to the exponents of the 51 mm CWV islands: $\alpha$ = 1.77 (0.05) and $\beta$ = 1.58 (0.04).
The perimeter dimension also agrees well with $D_l$ = 1.39 (0.01) for self-affine islands and $D_l$ = 1.35 (0.02) for the 51 mm CWV islands, 
whereas the agreement in volume dimension is not quite as good with 
$D_V$ = 2.41 (0.01) for self-affine islands and $D_V$ = 2.32 (0.03) for the CWV islands (Fig.\;\ref{fig:cluster_dimensions}a, b). 

However, we also see deviations of the CWV field from self-affine scaling. 
In particular, the power spectrum of CWV in hi-res has $\mu$ = 2.51 (0.39) as shown in Fig.\;\ref{fig:spectrum}, 
which would imply a larger value of $H$ $\approx$ 0.75 compared to the roughness of $H$ = 0.3 of self-affine surfaces 
that gives the best match for the CWV islands at 51 mm. 
The difference in power spectrum manifests in the differences in spatial patterns between hi-res CWV and the self-affine surface with $H$ = 0.3. 
Because the total variance is the same, hi-res CWV has less small-scale variability 
than the self-affine surface due to the steeper slope in the hi-res CWV power spectrum (compare Figs.\;\ref{fig:contours}a and b). 
Similarly, we find that $\alpha$ and $\beta$ vary differently as the threshold changes for hi-res CWV as compared to
the self-affine surface 
%(Fig.\;\ref{fig:CWV_exponents}a). 
(Fig.\;S1a). 
Thus, we speculate that precipitation dynamics may be decreasing $H$ for high values of the CWV threshold 
as compared to the appropriate $H$ for the bulk CWV field measured from the power spectrum. 
A more general scaling form than self-affine scaling may be
needed to capture all the statistical properties of the turbulent CWV field.

Although the CWV field is not exactly self-affine, islands on a self-affine
surface at $H$ = 0.3 do provide a good match to the CWV islands in hi-res at 51 mm
for all of the statistical properties we investigate in this study. 
Thus, in the next section, we connect analytical results based on 
self-affine scaling theory to the measured frequency distributions and fractal dimensions.

\subsection{Theoretical predictions of frequency distributions}\label{sec:theory_distribution}

Suppose that a series of evenly-spaced thresholds cuts through a self-affine topography 
and generates an ensemble of contour loops and the encircled islands at different levels (Fig.\;\ref{fig:contours}b).
The frequency distribution of the loop length in the contour ensemble is a 
power law whose slope is related to the roughness exponent, $H$ \citep{Kondev1995}. 
\citet{Pelletier1997} then showed that the frequency distribution of area within the contour loops also follows a power law as 
$\Pr(A) \propto A^{-\alpha}$, where $\Pr$ denotes frequency distribution, $A$ denotes loop area, and
\begin{equation}
     \alpha = 2 - \frac{H}{2}. 
    \label{eq:area_distribution}
\end{equation}
Eq.\;\eqref{eq:area_distribution} shows a reverse dependence of $\alpha$ on $H$, 
consistent with Fig.\;14 in \cite{Wood2011} which is based on a one-dimensional bounded cascade model for clouds.  
It's important to note that these power-law distributions of contour length and area apply to contours 
at all levels rather than at one particular threshold, and they also include the contours and areas of lakes within islands. 
For contours and islands at single levels near the mean level, 
Eq.\;\eqref{eq:area_distribution} still holds \citep{Rajabpour2009},
but when the level is raised far above the mean, Eq.\;\eqref{eq:area_distribution} overestimates $\alpha$ \citep{Olami1996}\footnote{\cite{Olami1996} neglected contours and areas associated with lakes within islands, whereas \cite{Rajabpour2009} considered all contours including contours within an island. We find that considering lakes inside islands reduces the bias in Eq.\;\eqref{eq:area_distribution} at thresholds close to the mean (0$\sigma$-1$\sigma$), but does not diminish the overall decreasing trend in $\alpha$ at high thresholds.}.

From Eq.\;\eqref{eq:area_distribution}, we derive a formula for the frequency distribution of island volume. 
The volume of an island scales as $V\propto A h$, 
where $A$ is the area and $h$ is the peak height of the island above the threshold. 
We assume that the area of lakes within the island is small compared to its total area, 
so that $A\propto R^2$ where $R$ is the island's radius. 
Define vertical width, $W(R)$, as the root-mean-square fluctuation of the surface height where the mean is taken over $R$. 
For a self-affine surface, it follows that $W^2(R) \propto R^{2H}$.
We further assume that the peak height of each island is proportional to the 
vertical width of the surface within the island's area coverage: $h \propto W(R) \propto R^H$, such that the volume scales as
\begin{equation}
 V \propto A R^H \propto R^{2+H}. 
    \label{eq:volume_scaling}
\end{equation}
Let $\Pr(V)$ be the frequency distribution of island volume, and assume that it has a power law form $\Pr(V) \propto V^{-\beta}$.
Substituting $\Pr(A) \propto A^{-\alpha}$ and Eq.\;\eqref{eq:volume_scaling} into $\Pr(A)\mathrm{d}A = \Pr(V)\mathrm{d}V$ yields 
\begin{equation}
\beta = \frac{2\alpha+H}{2+H}.
\end{equation}
Substituting for $\alpha$ using Eq.\;\eqref{eq:area_distribution} gives
\begin{equation}
    \beta = \frac{4}{2+H}. 
    \label{eq:volume_distribution}
\end{equation}
Therefore, the distributions of island area and volume both follow power laws for a self-affine topography, 
and the exponents of the power laws are controlled by the roughness exponent of the topography, $H$. 
Similar to $\alpha$, larger values of $H$ lead to smaller values of $\beta$, suggesting that both $\alpha$ and $\beta$
should follow similar trends when $H$ is varied.
Since Eq.\;\eqref{eq:area_distribution} overestimates $\alpha$ for a single threshold far above the mean, 
we expect Eq.\;\eqref{eq:volume_distribution} would also overestimate $\beta$ in that case since we have used 
Eq.\;\eqref{eq:area_distribution} in our derivation above.

The numerically generated self-affine surfaces in section \ref{sec:idealized_surface} 
suggest that self-affine surfaces with $H$ = 0.3 are an appropriate match to the CWV field for islands at the 51 mm convective threshold.
For this $H$ value, Eqs.\;\eqref{eq:area_distribution} and \eqref{eq:volume_distribution} predict that 
$\alpha$ = 1.85 and $\beta\approx$ 1.74, as compared to $\alpha$ = 1.78 (0.01) and $\beta$ = 1.60 (0.03) 
measured from the generated self-affine surfaces at 51 mm (2.0$\sigma$ above the mean). 
Thus, the theory correctly predicts that $\alpha$ is larger than $\beta$, 
but it over-predicts both values when applied to a single threshold high above the mean, 
consistent with previous work on $\alpha$ at different single thresholds \citep{Olami1996}.\footnote{The numerically generated self-affine surfaces give $\alpha$ = 1.84 for all contours at the mean threshold including lakes within islands (Fig.\;S3), and this value is in better agreement with the theoretical prediction of $\alpha = 1.85$. We do not report $\beta$ here because the volume is not well-defined for lakes within islands.}

The theoretical predictions for $\alpha$ and $\beta$ are related to each other via a scaling relation upon eliminating $H$ from Eqs.\;\eqref{eq:area_distribution} and \eqref{eq:volume_distribution}: 
\begin{equation}
    \alpha + \frac{2}{\beta} = 3. 
    \label{eq:scaling_relation}
\end{equation}
This relation allows the prediction of $\beta$ given $\alpha$ and vice versa without knowing the value of $H$.  Furthermore, for all $\alpha$ values between 1 and 2, $\beta$ is always smaller than $\alpha$ by Eq.\;\eqref{eq:scaling_relation}, which explains why $\beta$ is generally found to be smaller than $\alpha$ for precipitation clusters in prior works.
Despite the inaccuracies in the individual estimates of $\alpha$ and $\beta$, 
Eq.\;\eqref{eq:scaling_relation} holds well for observations and hi-res precipitation clusters (table \ref{tab:exponents}) 
and also for hi-res CWV islands and self-affine islands under a wide range of thresholds 
%(Fig.\;\ref{fig:CWV_exponents}b) 
(Fig.\;S1b) 
\footnote{Although we don't focus on GCM in the main text, it is interesting to note that Eq.\;\eqref{eq:scaling_relation} holds with $\alpha+2/\beta$ = 2.96 (0.14) for the very different values of $\alpha$ and $\beta$ that occur for GCM as compared to hi-res and observations ($\alpha$=1.10, $\beta$=1.07 as shown in Fig.\;\ref{fig:distributions_GCM}). }.

\subsection{Theoretical predictions of fractal dimensions}\label{sec:theory_dimensions}

For self-affine surfaces, the scaling theory also predicts the fractal dimension of contour loops, 
\begin{equation}
D_l = \frac{3 - H}{2},
\label{eq:perimeter_dimension}
\end{equation}
which was derived by \cite{Kondev1995} (partly based on a conjecture) and numerically confirmed by \citet{Rajabpour2009} and \citet{Nezhadhaghighi2011}.
Note that this dimension is the fractal dimension of a single contour loop, not the fractal dimension 
of all contours at the same level \citep[$D = 2 - H$ as in ][]{Mandelbrot1975}. 
For the volume fractal dimension, comparing its definition in Eq.\;\eqref{eq:D_V_definition} and the volume scaling in Eq.\;\eqref{eq:volume_scaling} gives that
\begin{equation}
D_V = 2 + H. 
\label{eq:volume_dimension}
\end{equation}

\begin{figure}[h!]
\centering
\noindent\includegraphics[width=1.0\linewidth]{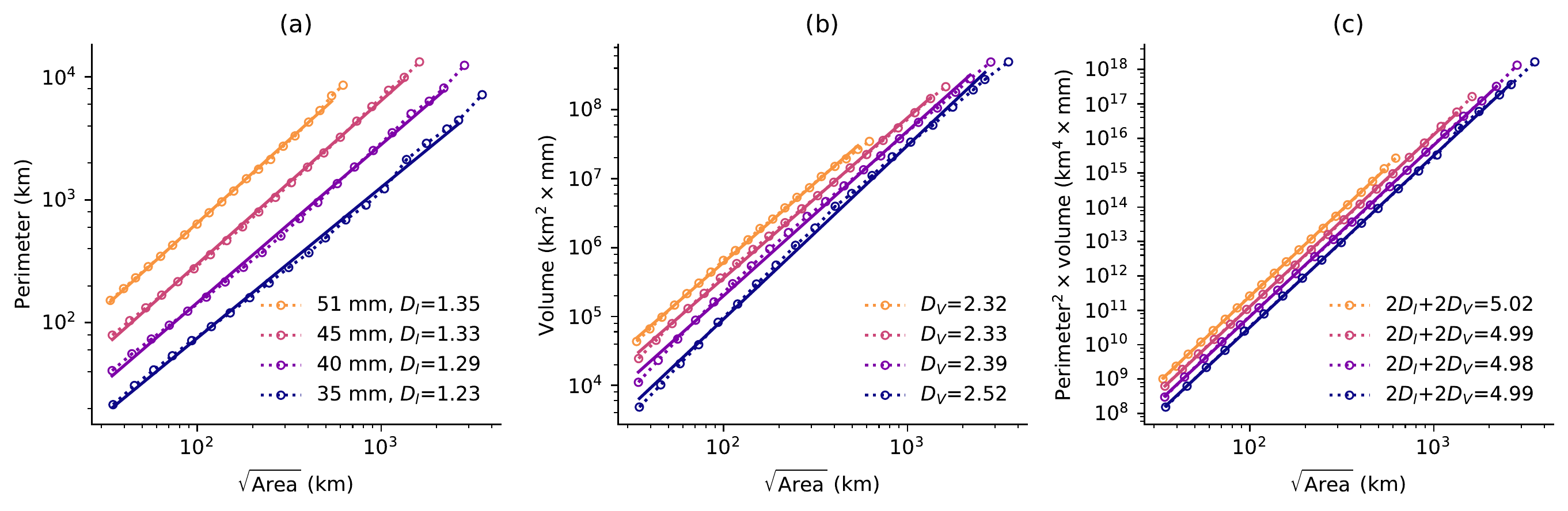}
\caption{Same as in Fig.\;\ref{fig:cluster_dimensions} but for hi-res CWV islands at thresholds of 51 mm, 45 mm, 40 mm, and 35 mm (from orange to navy, top to bottom). Starting from 51 mm, the scalings are consecutively shifted downwards by a factor of 2 for clarity. }
\label{fig:cluster_dimensions_CWV}
\end{figure}

For $H$ = 0.3, these theoretical predictions give $D_l$ = 1.35 and $D_V$=2.3. 
These values are in good agreement with the results for the self-affine surface with $H$ = 0.3 
which have $D_l$ = 1.39 (0.01) and $D_V$ = 2.41 (0.01)
and CWV islands at the threshold of 51 mm which have $D_l$ = 1.35 (0.02) 
and $D_V$ = 2.32 (0.03), shown in Fig.\;\ref{fig:cluster_dimensions}. 
Unlike for $\alpha$ and $\beta$, $D_l$ and $D_V$ for the islands on self-affine
surfaces do not vary strongly as the threshold is varied, but there is some 
evidence for systematic variations in hi-res CWV island dimensions
(Figs.\;\ref{fig:cluster_dimensions_CWV} and 
S2a).
%\ref{fig:CWV_dimensions}a).

Similar to the spirit of Eq.\eqref{eq:scaling_relation}, 
we can eliminate $H$ by combining Eqs.\eqref{eq:perimeter_dimension} and \eqref{eq:volume_dimension} and obtain 
\begin{equation}
2D_l + D_V = 5. 
\label{eq:dimension_relation}
\end{equation}
Eq.\;\eqref{eq:dimension_relation} holds approximately for the precipitation clusters in 
observations and hi-res (Table \ref{tab:exponents} and Fig.\;\ref{fig:cluster_dimensions}c). 
Note that Table \ref{tab:exponents} shows $2D_l + D_V$ based on individual $D_l$ and $D_V$ from different datasets, 
whereas Fig.\;\ref{fig:cluster_dimensions}(c) shows the scaling exponent measured from regressing $R^{2D_l + D_V} \sim l^2V$ in the log-log space. 
Eq.\;\eqref{eq:dimension_relation} also holds approximately for the self-affine 
surface and CWV islands at 51 mm (Table \ref{tab:exponents}) and 
also for a wide range of thresholds (Fig.\;\ref{fig:cluster_dimensions_CWV}c and 
%Fig.\;\ref{fig:CWV_dimensions}b).  
Fig.\;S2b).

Overall, the predictions based on the self-affine scaling theory provide considerable insight
into how the roughness of the CWV field controls the statistical properties of the CWV islands, 
even though there are some inaccuracies related to the intrinsic limitations in the theory 
(which overestimates $\alpha$ and $\beta$ for thresholds high above the mean)
and related to the deviation of the CWV field from self-affine scaling.

\section{Conclusions and discussion}\label{sec:conclusion}

We have shown from observations and a high-resolution simulation with explicit
convection that tropical precipitation clusters can be seen as islands on a
rough CWV topography cut by a convective threshold, analogous to the actual islands above sea level on Earth's relief.  
The physical basis for this link between precipitation clusters and CWV islands is the onset of precipitation at a
critical CWV level, which has been widely found in observations and simulations of the tropical atmosphere. 
Using the hi-res simulation as an idealized representation of the tropical atmosphere, we find that the CWV
islands at a convective threshold match precipitation clusters 
in the power-law frequency distributions of area and volume and also in their fractal dimensions.   
The frequency distributions of CWV island also follow power laws at a wide range of
other CWV thresholds, suggesting that the existence of power-law
distributions is not related to specific precipitation dynamics such as gust
fronts within the precipitation clusters, but is instead a general property of thresholded islands
on the CWV field.

We further assume that the CWV field is self-affine which allows us to apply the self-affine scaling theory. 
By numerically generating self-affine surfaces, we find that the CWV islands at the convective
threshold are well-matched by islands on a self-affine surface with a roughness
exponent of $H$ = 0.3 at the same threshold.  
Within the self-affine framework, the roughness exponent
of the topography governs the statistical properties of the islands. 
Previous work gave analytical expressions for the area distribution exponent
($\alpha$) and the perimeter fractal dimension ($D_l$). Here, we further
derive expressions for the volume distribution exponent ($\beta$) and 
the volume fractal dimension ($D_V$).  While the expressions for the fractal
dimensions are accurate, the expressions for $\alpha$ and $\beta$ are 
overestimates. The overestimation is likely due to the scaling theory being applicable
to all contours at all levels, not contours at the convective threshold which is high above the mean level. 

The roughness of idealized self-affine surfaces that gives the best correspondence to CWV islands ($H$ = 0.3)
is lower than the roughness directly measured from the CWV power spectrum ($H$ $\approx$ 0.75). 
We speculate that the roughness may effectively be lower in regions of precipitation, but it is also possible that the 
turbulent CWV field would be better described by a more general scaling (e.g., multifractals). 
Hence, deviations from the simple self-affine scaling in the CWV field should be investigated in future work.
Nonetheless, we derive a scaling relation from the scaling theory
that directly relates $\alpha$ to $\beta$, and a similar relation that connects $D_l$ and $D_V$. 
These scaling relations are approximately satisfied by the precipitation clusters and CWV islands across different thresholds. 
Given the discrepancies between the $H$-value best corresponding to 
CWV islands and the $H$-value measured from power spectra, 
these scaling relations are particularly useful as they don't involve $H$. 

The framework presented here connects precipitation clusters to the properties
of the CWV field, but the question of what determines the roughness of the CWV
field has not been addressed.  Horizontal diffusion and noise play important roles in
existing stochastic models of the CWV field \citep{Craig2013, Hottovy2015,
Ahmed2019}. In addition, horizontal advection by rotational winds (e.g., as in
two-dimensional turbulence) and gravity wave dynamics \citep{Stiassnie1991}
may also contribute to the scaling behavior of CWV.  One complication with
associating precipitation clusters with CWV islands is that precipitation
itself reduces the local volume of CWV islands, but this issue can be avoided
by considering the column moist static energy (CMSE) which is not affected by
condensation and precipitation. Under the weak-temperature-gradient
approximation \citep[e.g., ][]{Neelin1987}, the spatial patterns of water vapor and moist static energy are similar, 
and we expect CMSE islands to behave similarly to the CWV islands. 
The deviation of the CWV field from self-affinity is also worthy of further research. 
The distributions of CWV islands best-matching the distributions of precipitation clusters
are explained by self-affine surfaces with $H$ = 0.3, which is close to $H\simeq0.39$ as given by the KPZ universality class \citep{Pagnani2015}. 
Therefore, more work is needed to confirm whether tropical CWV displays KPZ-type behavior, 
and to identify the physical mechanism in precipitation dynamics that may give rise to the observed scaling relations. 
Such mechanism may be responsible for the smaller roughness exponent 
associated with the statistics of CWV islands at a high threshold, 
which is different from the larger $H$ value of the bulk CWV field as measured from its power spectrum.

An additional future avenue for research is to examine the response of precipitation cluster statistics to climate change
\citep[cf.][]{Quinn2017b}, particularly in high-resolution simulations that
have extensive power-law ranges. 
Eq.\;\eqref{eq:scaling_relation} suggests that any changes in the power-law exponent 
for the area distribution under warming would be directly related to 
changes in the exponent for the volume distribution, and thus affect the spatially integrated impacts of strong precipitation events. 

\iffalse
It would be interesting to see if there are
changes in the roughness of the CWV field as the climate changes and how these
relate to changes in the frequency distribution exponents and fractal
dimensions of precipitation clusters. 
\fi

\section*{Acknowledgements}
We acknowledge support from NSF AGS 1552195 and 1749986 and from the mTerra
Catalyst Fund.  
We thank David Neelin, Tom Beucler, Tim Cronin, William Boos, and Yi Ming for helpful discussions. 
We thank William Boos for providing the hi-res output, and we thank
Marat Khairoutdinov for making SAM available to the community.
We acknowledge high-performance computing support from 
Cheyenne (doi:10.5065/D6RX99HX) provided by NCAR's Computational and Information Systems Laboratory, 
sponsored by the National Science Foundation.

\section*{Conflict of interest}
The authors declare no conflict of interest. 

\clearpage

\appendix

% Add descriptions of figures
%\numberwithin{equation}{section}
%\numberwithin{figure}{section}
%\numberwithin{table}{section}

\section{Meanings of symbols}

\begin{table}[!h]
\caption{Meanings of symbols in the main text. }
\centering
\begin{tabular}{ll}
\headrow
\thead{Symbol}	&\thead{Meaning}\\
$\alpha$		&Cluster area exponent\\
$\beta   $		&Cluster volume exponent\\
$\sigma$		&Standard deviation of CWV\\
$\mu$			&Power spectrum exponent\\
$A$			&Cluster area\\
$C$			&Proportionality factor from CWV to precipitation\\
$D_l$			&Perimeter fractal dimension \\
$D_V$	 	&Volume fractal dimension \\
$H$	 		&Roughness (Hurst) exponent\\
$l$	 		&Cluster perimeter length\\
$P(\mathbf{r})$	&Precipitation at location $\mathbf{r}$\\
$\Pr(X)$		&Frequency distribution of $X$\\
$R$			&Cluster radius\\
$V$			&Cluster volume\\
\hline
\end{tabular}
\label{tab:exponent_meaning}
\end{table}

\section{Results for GCM and for CWV power spectra}\label{sec:GCM_power_spectrum}

\begin{figure}[h!]
\centering
\noindent\includegraphics[width=0.8\linewidth]{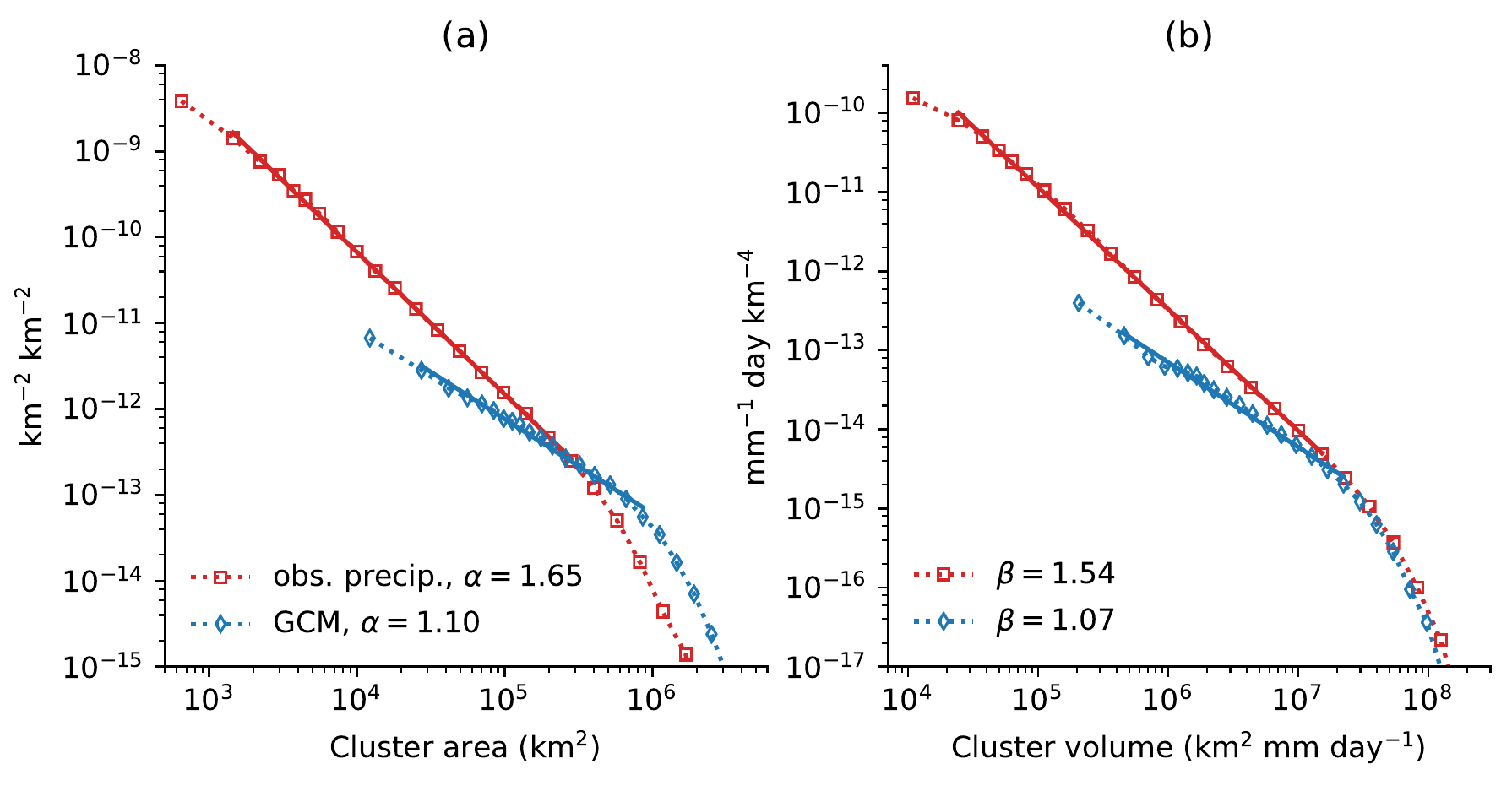}
\caption{The frequency distributions of (a) cluster area and (b) cluster volume 
for the GCM (blue diamonds) and observations (red squares). 
The selected region, regression method and normalization are the same as in Fig.\;\ref{fig:distributions}.}
\label{fig:distributions_GCM}
\end{figure}

In the GCM simulation, the precipitation cluster area and volume also follow power laws, 
but the exponents are shallower than those with the observations and hi-res simulation. 
As shown in Fig.\;\ref{fig:distributions_GCM}, the GCM simulation has 
$\alpha$ = 1.10 (0.07) for cluster area and $\beta$ = 1.07 (0.05) for cluster volume 
compared to $\alpha$ = 1.65 (0.04) and $\beta$ = 1.54 (0.04) in observations. 
This discrepancy remains if 6-hourly averaged precipitation is used for observations and hi-res 
to be consistent with the 6-hourly precipitation used for GCM. 

\begin{figure}[h!]
\centering
\noindent\includegraphics[width=0.4\linewidth]{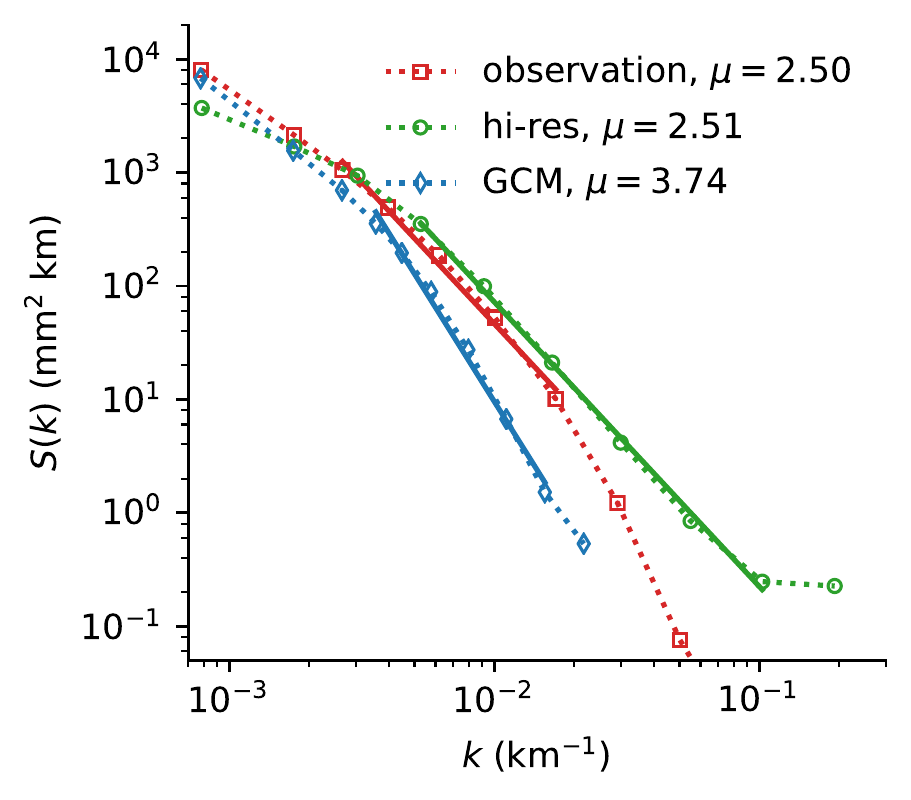}
\caption{One-dimensional power spectra of CWV as a function of wavenumber. 
The spectra are based on observations (red squares), hi-res simulation (green circles), 
and GCM simulation (blue diamonds) in the respective equatorial regions. 
The solid lines are linear regressions in the log-log space, and their extents correspond to the regression ranges. }
\label{fig:spectrum}
\end{figure}

One-dimensional spectra in the zonal direction of CWV for observations, hi-res, and GCM are shown in Fig.\;\ref{fig:spectrum}. 
The spectra are binned in the log wavenumber space with the bin widths rounded to multiples of the smallest wavenumber, 
$k_0 = 2\pi/L_x$, where $L_x$ is the domain width. 
The same tropical domains as in the main text are used to calculate the spectra, and 
the spectra are calculated at each latitude and then averaged in latitude and time.
We apply the Hann window in the zonal direction of the CWV fields of observations and GCM to reduce spectral leakage, 
and although not necessary, we also apply it in the case of hi-res for consistency. 
Similarly as for $\alpha$ and $\beta$, we measure the spectrum slope ($\mu$) by applying linear regression on the binned power spectrum. 
The regression ranges of the power spectra are matched to those of cluster area distributions as follows. 
$R$ is approximately related to $A$ by $R^2\approx 5A$ when averaged across all clusters for all datasets, 
% R^2\approx aA, a = 3.88 for TRMM, 5.79 for SAM, and 6.31 for CESM
and thus $\sqrt{A}$ corresponds to wavenumber $k = 2\pi/R \approx 2\pi/\sqrt{5A}$. We use this conversion between $A$ and $k$ to match the regression ranges, with the exception of the ERA5 CWV spectrum due to smoothing at small scales. 
%\texttt{why can't you use this approach for GCM? I'd recommend trying to use it for this too for consistency since any smoothing would affect both the clusters and the spectrum. Looks to me like the ranges you use for GCM do approximately match up but maybe I'm estimating wrong!} (It's actually the ERA5 CWV spectrum that we don't use this approach.)

For the hi-res simulation which has the best resolved CWV field of the three datasets, 
we find $\mu$ = 2.51 (0.39). The relation $\mu=2H+1$ then implies $H$ $\approx$ 0.75. 
Measurements of $H$ using the second-order structure function and 
detrended fluctuation analysis \citep{Bakke2007} for hi-res CWV yield somewhat smaller values for $H$ of 0.62 and 0.69, respectively. 
According to Eqs.\;\eqref{eq:area_distribution} and \eqref{eq:volume_distribution}, 
the self-affine scaling theory predicts that the steeper CWV power spectrum in GCM, $\mu$ = 3.74 (0.48), 
leads to a greater value of $H$ and thus smaller $\alpha$ and $\beta$ compared to observations and hi-res simulation. 
Indeed, the $\alpha$ and $\beta$ exponents for the GCM are much smaller than those for the observations (Fig. \ref{fig:distributions_GCM}).



\section*{Supplementary material}

# Supplementary figures for article "Tropical precipitation clusters as islands on a rough water-vapor topography"

**Ziwei Li**[1], **Paul A. O'Gorman**[1], **Daniel H. Rothman**[2]

[1] Department of Earth, Atmospheric and Planetary Science
Massachusetts Institute of Technology
Cambridge, MA, United States

[2] Lorenz Center, Department of Earth, Atmospheric, and Planetary Sciences
Massachusetts Institute of Technology
Cambridge, MA, United States

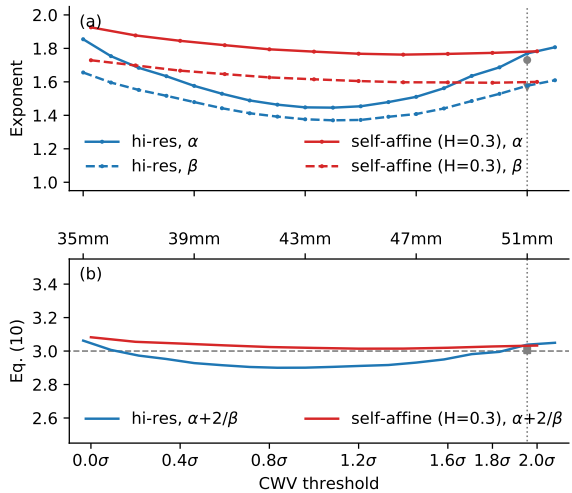

**Figure S1:** (a) Exponents of power-law frequency distributions and (b) scaling relation $\alpha + 2/\beta$ (Eq. 10) as functions of CWV threshold for hi-res CWV islands (blue) and islands on self-affine surfaces with $H = 0.3$ (red). In (a), the solid lines show the island area exponent ($\alpha$), and the dashed lines show the island volume exponent ($\beta$). The single circle, triangle, and square markers in grey show $\alpha$, $\beta$, and $\alpha + 2/\beta$, respectively, for hi-res precipitation clusters. The self-affine surfaces have the same mean and variance as the hi-res CWV field. The vertical dotted line shows the convective threshold of 51 mm.

576576

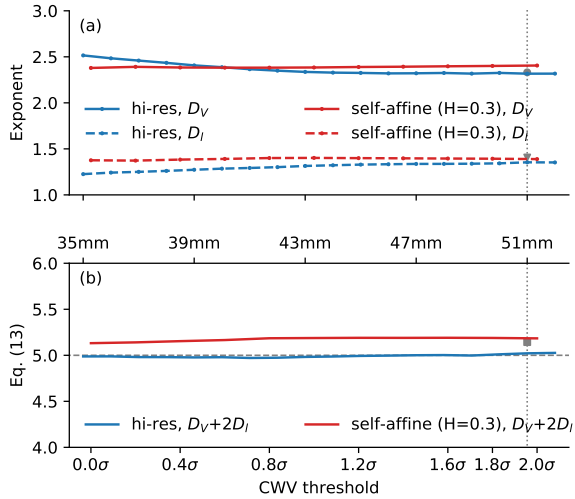

**Figure S2:** (a) Fractal dimensions and (b) their scaling relation $2D_l + D_V$ (Eq. 13) as functions of CWV threshold for hi-res CWV islands (blue) and islands on self-affine surfaces with $H$ = 0.3 (red). In (a), the solid lines show the cluster volume dimension ($D_V$), and the dashed lines show the cluster perimeter dimension ($D_l$). The single circle, triangle, and square markers in grey show $D_V$, $D_l$, and $2D_l+D_V$, respectively, for hi-res precipitation clusters. The self-affine surfaces have the same mean and variance as the hi-res CWV field. The vertical dotted line shows the convective threshold of 51 mm.

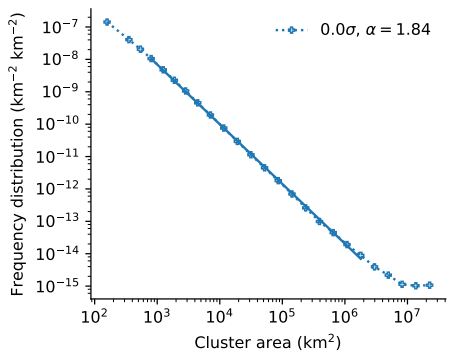

**Figure S3:** Frequency distribution of contour areas, including lakes within islands, at the mean threshold of the self-affine surfaces with $H$ = 0.3. The solid line represents linear regression in log-log space.

2



\end{document}